# Ultrafast optical control of magnetization in EuO thin films

T. Makino[1,*], F. Liu[2,3], T. Yamasaki[4], Y. Kozuka[2], K. Ueno[5,6], A. Tsukazaki[2], T. Fukumura[6,7], Y. Kong[3], and M. Kawasaki[1,2]

[1] *Correlated Electron Research Group (CERG) and Cross-Correlated Materials Research Group (CMRG), RIKEN Advanced Science Institute, Wako 351-0198, Japan*

[2] *Quantum Phase Electronics Center and Department of Applied Physics, University of Tokyo, Tokyo 113-8656, Japan,*

[3] *School of Physics, Nankai University, Tianjin 300071, China*

[4] *Institute for Materials Research, Tohoku University, Sendai, 980-8577, Japan*

[5] *Graduate School of Arts and Sciences, University of Tokyo, Tokyo 153-8902, Japan*

[6] *PRESTO, Japan Science and Technology Agency, Tokyo 102-0075, Japan,*

[7] *Department of Chemistry, University of Tokyo, Tokyo 113-0033, Japan,*

All-optical pump-probe detection of magnetization precession has been performed for ferromagnetic EuO thin films at 10 K. We demonstrate that the circularly-polarized light can be used to control the magnetization precession on an ultrafast time scale. This takes place within the 100 fs duration of a single laser pulse, through combined contribution from two nonthermal photomagnetic effects, i.e., enhancement of the magnetization and an inverse Faraday effect. From the magnetic field dependences of the frequency and the Gilbert damping parameter, the intrinsic Gilbert damping coefficient is evaluated to be $\alpha \approx 3\times10^{-3}$.

Optical control of the spin in magnetic materials has been one of the major issues in the field of spintronics, magnetic storage technology, and quantum computing[1]. One type of the spin controls is based on the directional manipulation in the spin moments[2]. This yields in observations of spin precession (reorientation) in antiferromagnets and ferromagnets when magnetization is canted with respect to an external field[3–14]. In many previous reports, the spin precession has been driven with the thermal demagnetization induced with the photo-irradiation. Far more intriguing is the ultrafast nonthermal control of magnetization by light[8,10,14], which involves triggering and suppression of the precession. The precession-related anisotropy is expected to be manipulated through laser-induced modulation of electronic state because the anisotropy field originates from the magnetorcrystalline anisotropy based on the spin-orbit coupling. Recently, the spin precession with the non-thermal origin has been observed in bilayer manganites due to a hole-concentration-dependent anisotropic field in competing magnetic phases[15]. Despite the success in triggering the reorientation by ultrafast laser pulses, the authors have not demonstrated the possibility of the precessional stoppage.

On the other hand, photomagnetic switch of the precession has been reported in *ferrimagnetic* garnets with use of helicity in light[8,10]. The authors attributed the switching behavior to long-lived photo-induced modification of the magnetocrystalline anisotropy[16] combined with the inverse Faraday effects[17,18]. The underlying mechanism for the former photo-induced effect is believed to be redistribution in doped ions[16]. This is too unique and material-dependent, which is not observed in wide variety of magnets. For establishing the universal scheme of such "helicity-controllable" precession, it should be more useful to rely on more generalized mechanisms such as the carrier-induced ferromagnetism and the magnetic polarons[19]. A ferromagnet should be a better choice than a ferrimagnet or an antiferromagnet, *e.g.*, for aiming a larger-amplitude modulation by making use of its larger polarization-rotation angle *per unit length*. We have recently reported the optically-induced enhancement of magnetization in ferromagnetic EuO associated with the optical transition from the 4*f* to 5*d* states[20]. This enhancement was attributed to the strengthened collective magnetic ordering, mediated with the magnetic polarons. The helicity-controllable precession is expected to be observed in EuO by combining the photo-induced magnetization enhancement[20] with the inverse Faraday effect[17,18] because the magnetization is related to the magnetic anisotropy. The occurrence of the inverse Faraday effects is expected because of the high crystalline symmetry in EuO[17,18]. The magnetic properties of EuO are represented by the saturation magnetization of 6.9 $\mu_B$/Eu, the Curie temperature of 69 K, and the strong in-plane anisotropy[21,22].

In this article, we report observation of the photomagnetic switch of the spin precession with the nonthermal origin in a EuO *thin film* for the first time to the best of our knowledge. Due to the above-mentioned reasons, our findings deserve the detailed investigations such as the dependence on the circularly polarized lights, the frequency of precession, the Gilbert damping constants, and the magnitudes of the photo-induced anisotropic field.

EuO films were deposited on YAlO$_3$ substrate using a pulsed laser deposition system with a base pressure lower than $8\times10^{-10}$ Torr[22]. The EuO films were then capped with AlO$_x$ films *in-situ*. EuO and AlO$_x$ layers have thicknesses of 310 and 30 nm, respectively. The film turned out to be too insulating to be quantified by a conventional transport measurement method. The all-optical experiments have been performed using a standard optical set-up with a Ti:sapphire laser combined with a regenerative amplifier (accompanied with optical parametric amplifier). The wavelength, width, and repetition rate of the output pulse were 650 nm, ≈100 fs, and 1 kHz, respectively. The pump and probe pulses were both incident on the film at angles of $\theta_H \approx 45$ degree from the direction normal to the film plane as shown in inset of Fig. 1. The direction of the probe beam is slightly deviated from that of the pump so as to ensure the sufficient spatial separation of the reflected beams. The angle between the sample plane and the external field is approximately 45 degree. The polarization rotation of the reflected probe pulses due to the Kerr effect was detected using a Wollaston prism and a balanced photo-receiver. The pump fluence was approximately 0.5 mJ/cm$^2$. A magnetic field was applied using a superconducting electromagnet cryostat. The maximum applied magnetic field was $\mu_0 H \approx 3$ T. All the measurements were performed at 10 K.

Figure 1 shows a magneto-optical Kerr signal as a function of the pump-probe delay time for a EuO film at $\mu_0 H = 3.2$ T under the irradiation of right-circularly polarized (σ⁺) light. Its time trace is composed of instantaneous increase and decay of the Kerr rotation, and superimposed oscillation[20]. The oscillatory structure corresponds to the precession of magnetization. A solid (black) curve in Fig. 1 shows the result of fit to the experimental data using an exponentially decaying function and a damped oscillatory function. The precession is observed even with the linearly polarized light, which is consistent with the fact that EuO is a ferromagnet at this temperature.

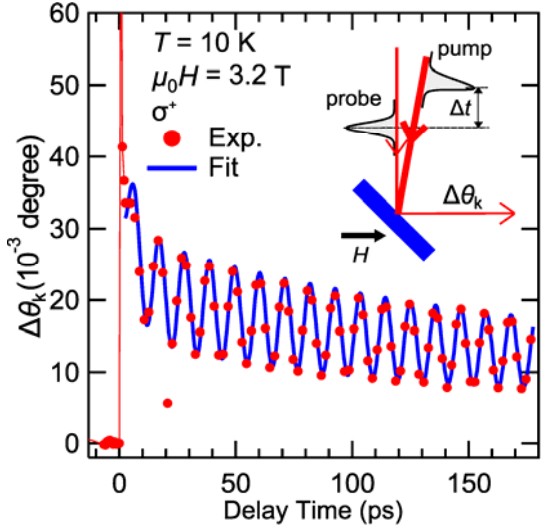

FIG. 1 (color online). Time-resolved Kerr signals recorded for a EuO thin film at a magnetic field of 3.2 T, and a temperature of 10 K for right circularly-polarized (σ⁺) light. The inset schematically shows the experimental arrangement. Experimental data are shown by (red) symbols, while the result of fit was shown by a full (blue) line.

For the detailed discussion of the precession properties, we subtracted the non-oscillatory part from the Kerr signal as a background. The results are shown in Fig. 2 for nine magnetic fields and for σ⁺ and left-circularly polarization (σ⁻). The subtracted data were then fitted with the damped harmonic function in the form of $A\exp(-t/\tau) \sin(2\pi f t+\varphi)$, where $A$ and $\varphi$ are the amplitude and the phase of oscillation, respectively. The amplitude of the precession was not found to depend on the plane of the linear polarization of the pump pulse. There is a linear relationship between the amplitude of precession and the pump fluence for the excitation intensity range measured. It is also noticed in Fig. 2 that the precession amplitudes are different each other for the two helicities (σ⁺ and σ⁻) even at the same magnetic fields. The magnetic field dependence of the amplitude is summarized in Fig. 3(d).

The minimum precession amplitude appears at around $\mu_0 H$ = +0.4 T for the σ⁻, while the minimum is observed at $\mu_0 H$ = −0.4 T for the σ⁺ as indicated by the shaded regions. To explain such disappearance of the precession and the triggering of precession even with a linearly-polarized light, it is necessary to take two effects into account. One of the effects that we seek should be odd with respect to the helicity of light. An effective magnetic field through the inverse Faraday effect is plausible to interpret this phenomenon because this satisfies the above requirements [$\mathbf{H}^F_{//}$ (black arrows) in Figs. 3(a) and 3(b)]. While the normal Faraday effect causes difference in the refractive indices for the left and right circularly polarized lights propagating in a magnetized medium, it is also possible to induce the inverse process where circularly polarized lights create a magnetization or an effective field[17,18]. The field associated with the inverse Faraday effect changes its sign when the circular polarization is changed from left-handed to right-handed.

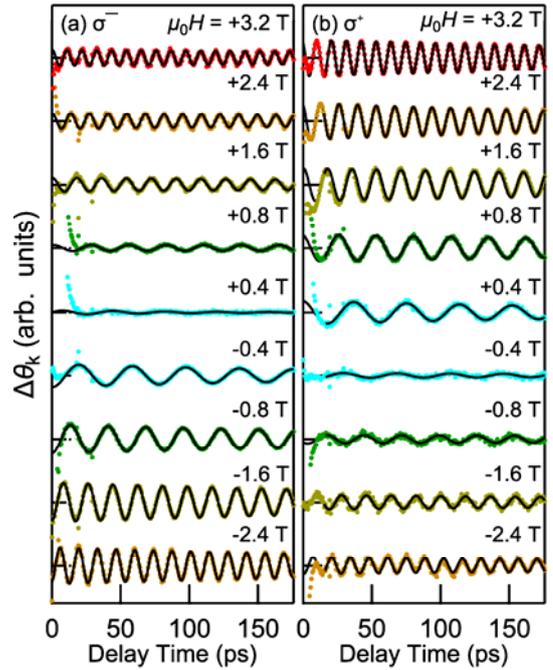

FIG. 2 (color online). A series of precession signals under various magnetic fields for right- and left-circularly polarized (σ⁺ and σ⁻) lights. Solid circles show the experimental data for which the non-oscillatory background is subtracted, while solid curves represent the calculated data as described in the text.

The other effect involved is considered to be the photoinduced enhancement of the anisotropic field (magnetization) associated with the $4f \rightarrow 5d$ optical transition [$\Delta\mathbf{M}$ (purple arrows) in Figs. 3(a) and 3(b)][20]. Our previous work quantified the photoinduced enhancement of the magnetization to be $\Delta M/M \approx 0.1\%$[20]. The amplitude of precession is determined from combination of $\Delta \mathbf{M}$ with the component of the inverse-Faraday field ($\mathbf{H}^F_{//}$) approximately projected onto the easy-axis direction. For example, no precession is triggered for $\mu_0 H$ of +0.4 T (−0.4 T) and $\sigma^-$ ($\sigma^+$), which is due to the balance of these two effects [Fig. 3(a)]. On the other hand, constructive contribution of these effects leads to a change in the direction of the magnetization [two dashed lines and a red arrow in Fig. 3(b)], which enhances the precession amplitude. The strength of the photoinduced field $H^F$ can be estimated to be approximately 0.2 T at the laser fluence of 0.5 mJ/cm². The derivation was based on Eq. (17) of Ref. 10. For more quantitative discussion for the suppression and enhancement of precession, the effect of the perpendicular component of inverse Faraday field is necessary to be taken into account. Such analysis is not performed here because this goes beyond the scope of our work.

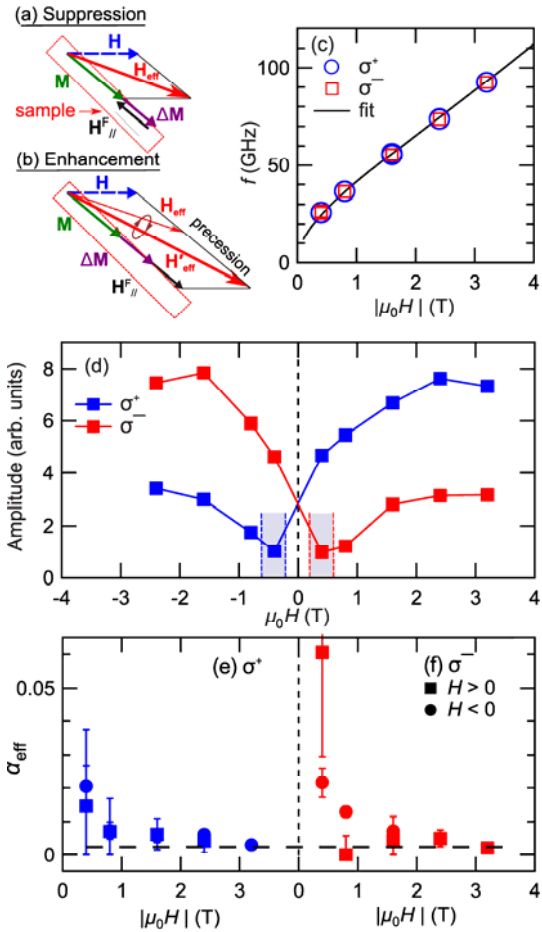

FIG. 3 (color online). Graphical illustrations of the magnetic precession; its suppression (a) and enhancement (b). **M** is a magnetization (green), **H** the external magnetic field (blue), **H**$_{\text{eff}}$ the effective magnetic field (red), $\Delta$**M** a photo-induced magnetization enhancement (purple), and the **H**$^F_{//}$ the inverse Faraday field (black). The situations of suppression correspond to the conditions of 0.4 T for $\sigma^-$ and −0.4 T for $\sigma^+$. The situations of enhancement are for opposite cases. Magnetic field dependences of the magnetization precession related quantities for $\sigma^+$ and $\sigma^-$; precession frequency $f$ (c), amplitude (d), and effective Gilbert damping $\alpha_{\text{eff}}$ (e) (f).

For the derivation of the precession-related parameters, we plot the frequency ($f$) and the amplitude of the magnetization precession for two different helicities as a function of $H$ with closed symbols in Figs. 3(c) and 3(d). To deduce the Landé g-factor g, we calculated $f(H)$ using a set of Kittel equations for taking the effect of tilted geometry into account as[12,23]:

$$f = \gamma \sqrt{H_1 H_2} \quad (1)$$

$$H_1 = H\cos(\theta_H - \theta) - M_{\text{eff}} \cos^2\theta \quad (2)$$

$$H_2 = H\cos(\theta_H - \theta) - M_{\text{eff}} \cos 2\theta \quad (3)$$

Here, $\gamma$ is the gyromagnetic ratio ($g\mu_B/h$), $\mu_B$ the Bohr magneton, $h$ Planck's constant, and $\theta_H$ an angle between the magnetic field and direction normal to the plane. $M_{\text{eff}}$ is the effective demagnetizing field given as $M_{\text{eff}} = M_S - 2K_\perp/M_S$, where $M_S$ is the saturation magnetization and $K_\perp$ is the perpendicular magnetic anisotropy constant. $\theta$ is an equilibrium angle for the magnetization, which obeys the following equation:

$$\sin 2\theta = (2H/M_{\text{eff}}) \sin(\theta - \theta_H) \quad (4)$$

A solid (black) line in Fig. 3(c) corresponds to the result of the least-square fit for the frequency $f$. The values of parameters are $g \approx 2$ and $\mu_0 M_{\text{eff}} \approx 2.4$ T. The $g$ value is consistent with the one derived from the static ferromagnetic resonance measurement[24].

Having evaluated the precession-related parameters such as $g$ and $M_{\text{eff}}$, we next discuss $H$ dependence of an effective Gilbert damping parameter $\alpha_{\text{eff}}$. This quantity is defined as:

$$\alpha_{\text{eff}} = \frac{1}{2\pi f \tau} \quad (5)$$

Figures 3(e) and 3(f) show the effective Gilbert damping parameter $\alpha_{\text{eff}}$ derived from the decay time constant ($\tau$) for $\sigma^+$ and $\sigma^-$, respectively. Despite relatively strong ambiguity shown with bars in Figs. 3(e) and 3(f), the damping parameters $\alpha_{\text{eff}}$ is not independent of the magnetic field. It is rather appropriate to interpret that for $\alpha_{eff}$ for low fields are larger than those at higher fields. Such dependence on magnetic field is consistent with those in general observed for a wide range of the ferrimagnets and ferromagnets. Two-magnon scattering has been adopted for the explanation of this trend[25]. When the magnitude or direction of the magnetic anisotropy fluctuates microscopically, magnons can couple more efficiently to the precessional motion[25]. Such may cause an additional channel of relaxation. Due to the suppressed influence of the abovementioned two-magnon scattering, the higher-field data correspond to an intrinsic Gilbert damping constant $\alpha \approx 3\times10^{-3}$, as shown with a dashed (black) line in Figs. 3(e) and 3(f). This value is comparable with that reported in Fe[26,27,28,29] and significantly larger than that of yttrium iron garnet, which is known for intrinsically low magnetic damping[8,10,14].

In conclusion, we have reported the observation of magnetization precession and the dependence on light helicity in ferromagnetic EuO films. We attribute it to the photo-induced magnetization enhancement combined with the inverse Faraday effect. The magnetic field dependence of the precession properties allowed us the evaluation of the Gilbert damping constant to be $\approx 3\times10^{-3}$.

Acknowledgements—the authors thank K. Katayama, M. Ichimiya, and Y. Takagi for helpful discussion. This research is granted by the Japan Society for the Promotion of Science (JSPS) through the "Funding Program for World-Leading Innovative R&D on Science and Technology (FIRST Program)," initiated by the Council for Science and Technology Policy (CSTP) and in part supported by KAKENHI (Grant Nos. 23104702 and 24540337) from MEXT, Japan (T. M.).